\documentclass[journal]{IEEEtran}
\ifCLASSINFOpdf
\else
\fi
\usepackage{graphicx}		
\usepackage{amsmath}
\usepackage{widetext}
\usepackage{amssymb}
\usepackage{amsfonts}
\usepackage{color}

\begin{document}
%
\title{Multi-node optical frequency dissemination with post automatic phase correction}
%
%
%

\author{Liang Hu, Xueyang Tian, Guiling Wu,~\IEEEmembership{Member,~IEEE,} Mengya Kong, Jianguo Shen, and Jianping Chen
\thanks{Manuscript received xxx xxx, xxx; revised xxx xxx, xxx. This work was supported in part by the National Natural Science Foundation of China (NSFC) (61627871, 61535006, 61905143) and in part by Zhejiang provincial Nature Science Foundation of China (LY17F050003). (\textit{Corresponding author: Liang Hu (liang.hu@sjtu.edu.cn) and Guiling Wu (wuguiling@sjtu.edu.cn)})}
\thanks{L. Hu, X. Tian, G. Wu, M. K, J. Chen are with the State Key Laboratory of Advanced Optical Communication Systems and Networks, Department of Electronic Engineering, Shanghai Jiao Tong University, Shanghai 200240, China, and also with  Shanghai Institute for Advanced Communication and Data Science, Shanghai Jiao Tong University, Shanghai 200240, China (e-mail: liang.hu@sjtu.edu.cn; txy0220@sjtu.edu.cn; wuguiling@sjtu.edu.cn; kmy-mandy@sjtu.edu.cn; jpchen62@sjtu.edu.cn).}
\thanks{J. Shen is with the College of Physics and electronic information Engineering, Zhejiang Normal University, Jinhua, 321004, China (e-mail: shenjianguo@zjnu.cn).}
}

%
%

\markboth{Journal of Lightwave Technology,,~Vol.~xxx, No.~xxx, December~2019}%
{Shell \MakeLowercase{\textit{et al.}}: Bare Demo of IEEEtran.cls for IEEE Journals}
%



\maketitle

\begin{abstract}
We report a technique for coherence transfer of laser light through a fiber link, where the optical phase noise induced by environmental perturbation via the fiber link is compensated by remote users with passive phase noise correction, rather than at the input as is conventional. Neither phase discrimination nor active phase tracking is required due to the open-loop design, mitigating some technical problems such as the limited compensation speed and the finite compensation precision as conventional active phase noise cancellation.   We theoretically analyze and experimentally demonstrate that the delay effect introduced residual fiber phase noise after noise compensation is a factor of 7 higher than the conventional techniques. Using this technique, we demonstrate the transfer laser light through a 145-km-long, lab-based spooled fiber. After being compensated, the relative frequency instability in terms of overlapping Allan deviation is $1.9\times10^{-15}$ at 1s averaging time and scales down $5.3\times10^{-19}$ at 10,000 s averaging time. The frequency uncertainty of the light after transferring through the fiber relative to that of the input light is $(-0.36\pm2.6)\times10^{-18}$. As the transmitted optical signal remains unaltered until it reaches the remote sites, it can be transmitted simultaneously to multiple remote sites on an arbitrarily complex fiber network, paving a way to develop a multi-node optical frequency dissemination system with post automatic phase noise correction for a number of end users. 
\end{abstract}


%
\IEEEpeerreviewmaketitle

\section{Introduction}

\IEEEPARstart{O}{ptical} frequency references and clocks have achieved an unprecedented accuracy of better than 1 part in $10^{18}$, with an instability near 1 part in $10^{19}$ \cite{ludlow2015optical, poli2014optical, PhysRevLett.120.103201, Schioppo:2016aa, McGrew:2018aa}. Newly developed optical frequency dissemination schemes have attracted widespread research interest \cite{Giorgetta:2013aa, Riehle:2017aa}. Fiber-based optical frequency has recognised as an ideal solution for ultra-long haul dissemination because of fiber-optic's unique advantages of broad bandwidth, low loss, and high immunity to environmental perturbations, etc \cite{ma1994delivering, Riehle:2017aa}. Thanks to the rapid development over the last decades, optical frequency dissemination with even increasing performance and fiber lengths are expected to play a crucial role for science and technology including time and frequency metrology \cite{foreman2007coherent, daussy2005long, droste2013optical2, calonico2014high}, navigation, radio astronomy \cite{clivati2017vlbi, wang2015square} and tests of fundamental physics \cite{lisdat2016clock, hu2017atom}. To date, efforts have mainly focused on long-distance connections between just two locations connected by an optical fiber by correcting phase perturbations between the local and remote end \cite{ma1994delivering}, For example, frequency comparison between two separated Sr optical clocks located at PTB (Germany) and SYRTE (France) has been demonstrated with a 1415-km-long optical fiber link \cite{lisdat2016clock} and the gravity potential difference between the middle of a mountain and a location 90 km away  has been determined with a transportable Sr optical lattice clock from PTB \cite{grotti2018geodesy}.

 Schemes with a variety of different topological structures that are intended to meet various application requirements, ranging from point-to-point to multi-access, cascade, and star-shaped structures, have been experimentally demonstrated.  Delivering a stable optical frequency to multiple locations has been demonstrated by extracting a portion of the forward and backward transferred optical signals at any position along the fiber link as first proposed by Grosche \textit{et al.} \cite{grosche2014eavesdropping}. In this approach, many places along the fiber link can be served as an optical reference frequency and only one single fiber link stabilisation control loop is necessary. A clear disadvantage of this technique is that in case of malfunction of the link stabilisation system, all the access nodes will also be failure \cite{grosche2014eavesdropping, bai2013fiber, bercy2014line}. To overcome the above main drawback, ultrastable optical frequency dissemination schemes on a star topology optical fiber network have been proposed and experimentally demonstrated \cite{schediwy2013high, wu2016coherence}. Using this method, a highly synchronized the optical signal itself can be recovered at any remote sites by actively compensating the phase noise of each fiber link at each user end \cite{schediwy2013high, wu2016coherence}. 


\begin{figure*}[htbp]
\centering
\includegraphics[width=0.98\linewidth]{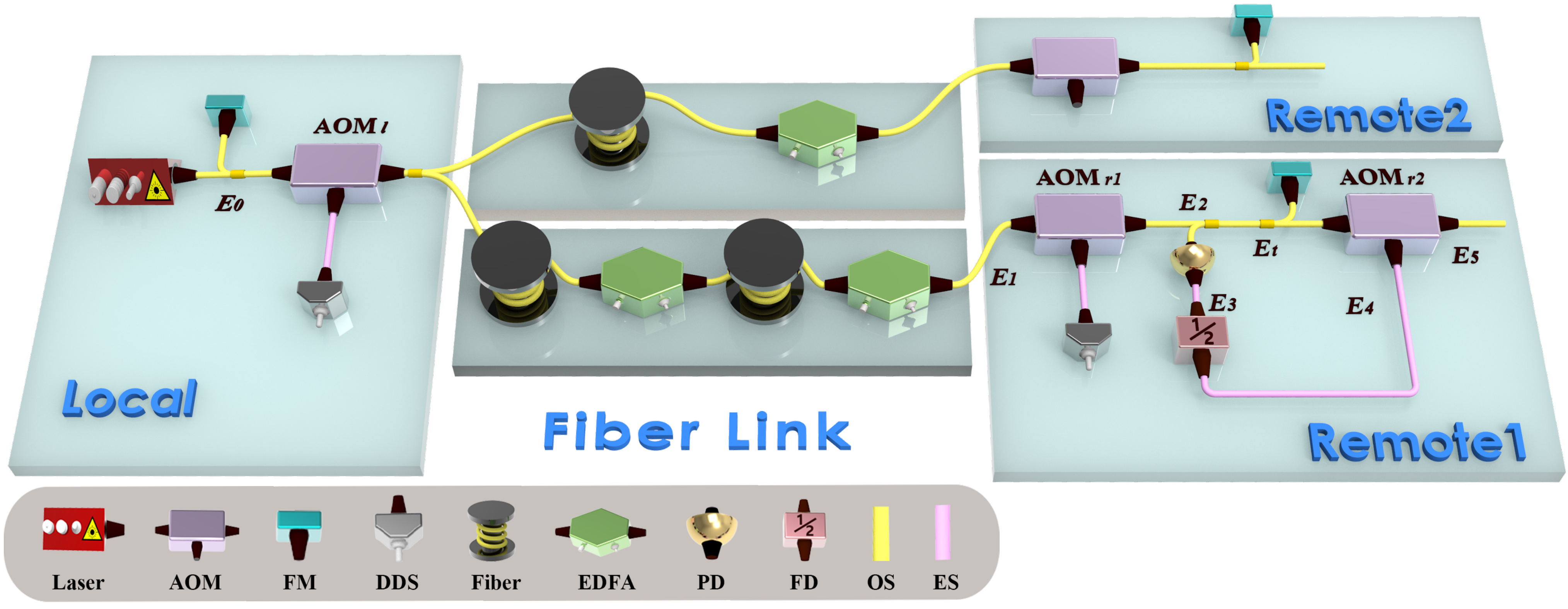}
\caption{Schematic view of our multi-node optical frequency transfer over a star topology fiber network.  The optical phase introduced by environment perturbations on the fiber links is passively compensated at the remote sites. FM: Faraday mirror, DDS: direct digital synthesis, EDFA: Erbium-doped fiber amplifier, AOM: acousto-optic modulator, PD: photo-detector, FD: frequency divider, OS: optical signal, ES: electrical signal.}
\label{fig1}
\end{figure*}


Optical phase noises of the transmitted frequency introduced by thermal and acoustic fluctuations of the fiber link can be actively cancelled either by physically modulating the fiber length or by shifting the carrier frequency via a voltage-controlled oscillator (VCO).  The main drawback related to the active approach is the need of complex circuits  to extract the phase error and drive the devices for phase correction in real time e.g., frequency divider with a high ratio to avoid the cycle slips. Although the method based on VCOs principally has an endless compensation range, it is hard to achieve precise compensation because of the high voltage sensitivity of the VCOs.  Additionally, the pulsing of the clock laser light and optical frequency transfer over free-space where the interruptions frequently happened call for short settling times of stabilization locks, namely the fast compensation speed, which is great challenge even for dedicated servo amplifiers\cite{falke2012delivering, gozzard2018stabilized}. 




Passive phase correction schemes based on frequency multiplexing and mixing to realize stable radio frequency (RF) dissemination via optical fiber have been proposed and experimentally  demonstrated \cite{pan2016passive, he2013stable, huang2016fiber, li2014phase, yu2014stable}. In comparison with the active schemes, the passive phase correction schemes would have a simple structure, a fast compensation speed, and an infinite compensation precision. In addition, the passive phase correction method can be achieved by both pre-phase cancellation in the transmitter and post-phase correction in the receiver \cite{pan2016passive, wu2013stable}. While RF frequency transfer based on passive phase noise cancellation has been experimentally demonstrated, the frequency multiplexing and dividing method is not suitable for optical frequency dissemination, in particular, for fiber based optical frequency dissemination.  

In this manuscript, we propose a technique for the coherence transfer of laser light through a fiber link, where the optical phase noise induced by environmental perturbation via the fiber link is automatically compensated by remote users with passive phase noise cancellation. Theoretical analysis demonstrates that the delay effect introduced residual fiber phase noise after noise compensation is a factor of 7 higher than the conventional techniques. The theoretical result was compared to experimental results that were obtained for optical frequency transfer over a 145 km fiber link. The experimental results confirm that the proposed scheme with post automatic phase correction will slightly degrade the performance of the system and, however, a state-of-art result can still be obtained.

The article is organized as follows. We illustrate the concept of multi-node optical frequency dissemination with post automatic phase noise cancellation in Sec. II and explain in Sec. III how to numerically analyze the delay limited phase noise power spectral density (PSD) based on the triple-pass technique for the first time. In Sec. IV, we discuss the experimental apparatus and results.  Finally, we conclude in Sec. V by briefly summarizing our results and providing an outlook. 



\section{Concept of multi-node optical frequency dissemination}
Figure \ref{fig1} shows coherent transfer of optical frequency based on a triple-pass technique with post automatic phase noise cancellation \cite{schediwy2013high, wu2016coherence, ma2018delay}. The laser light is coupled into a star topology fiber network. The transferred light can be extracted by each remote user. The Faraday mirrors (FMs) are separately installed at the local end and each remote end to rotate the polarization of light by 45$^{\circ}$ for a single pass and reflect the transferred light back. To distinguish the reflected light by the FM from the stray reflected light along the fiber network, the acousto-optic modulator (AOM$_l$) and AOM$_{r1}$ with different driving frequencies are inserted at the local and remote ends, respectively. For each remote user, the driving frequency of AOM$_{r1}$ is unique to discriminate its frequency, and additional AOM (AOM$_{r2}$) is cascaded after just AOM$_{r1}$ to correct the phase noise of the single-pass light. The extraction and compensation of the fiber phase noise are both carried out at the remote end. This is the fundamental difference between the double-pass scheme and the triple-pass scheme as discussed in \cite{schediwy2013high, wu2016coherence}. The beatnote is obtained on a photo-detector (PD) by heterodyne beating the single-pass light against the triple-pass light, acting as a phase detector that the phase noise introduced by each fiber link.  The different frequency of the light source enables remote users at different sites to compensate for the fiber phase noise independently. More importantly, using passive phase noise correction at each remote user enables significantly the improvement of the robustness of the system.  

Figure \ref{fig1} shows the diagram of the experimental setup for multi-node optical frequency dissemination through the fiber links. The electric field of the light from a narrow-linewidth laser is 
\begin{equation}
E_{0}\propto\cos(\nu t+\phi_s)
\end{equation}
where $\nu$ and $\phi_s$ are the angular frequency and the phase of the light, and here and in the following text we ignore the amplitude for conciseness. The light is passed through a local AOM (AOM$_l$) working at the angular frequency of $+\omega_l$ with the initial phase of $\phi_l$ and then coupled into a star topology fiber network. At the remote, taking the remote user 1 as an example, the fiber output light can be expressed as
\begin{equation}
E_{1}\propto\cos((\nu +\omega_l)t+\phi_s+\phi_l+\phi_{p})
\end{equation}
where $\phi_{p}$ is the added phase noise after transferring through the fiber. The received signal at the remote user 1  is frequency shifted by $+\omega_{r1}$ with an initial phase of $\phi_{r1}$ with AOM$_{r1}$. The electric field of the one-way is 
\begin{equation}
E_{2}\propto\cos((\nu+\omega_l+\omega_{r1})t+\phi_s+\phi_l+\phi_p+\phi_{r1})
\end{equation}

Assuming that the optical phase noise introduced by environmental perturbations on the fiber link in either direction is equal, the electric field of the light after transferring through the fiber between the local site and the remote for another roundtrip is 
\begin{equation}
E_{t}\propto\cos((\nu+3(\omega_l+\omega_{r1}))t+\phi_s+3(\phi_l+\phi_p+\phi_{r1}))
\end{equation}

$E_{t}$ heterodyne beats against $E_2$ on PD, the upper side of the heterodyne beatnote is 
\begin{equation}
E_{3}\propto\cos(2(\omega_l+\omega_{r1})t+2(\phi_l+\phi_p+\phi_{r1}))
\end{equation}


Afterwards we divide the angular frequency of $E_3$ with a factor of 2, resulting in 
\begin{equation}
E_{4}\propto\cos((\omega_l+\omega_{r1})t+(\phi_l+\phi_p+\phi_{r1}))
\end{equation}

To compensate the phase noise introduced by the fiber link, we cascade one more AOM (AOM$_{r2}$) working at the downshifted mode just after AOM$_{r1}$. Then we feed $E_{4}$ to the RF port of AOM$_{r2}$. After being compensated, the single-way light at the remote site is expressed as 
\begin{equation}
\begin{split}
E_5&\propto\cos(\nu t+\phi_s)+\cos\left[(\nu+2(\omega_l+\omega_{r1}))t\right.\\
&\,\,\,\,\,\,\,\,\,\,\,\,\,\,\,\,\,\,\,\,\,\,\,\,\,\,\,\,\,\,\left.+\phi_s+2(\phi_l+\phi_p+\phi_{r1})\right]+\text{others}
\end{split}
\label{eq7}
\end{equation}
where others represent multiple-trip optical signals between the local site and the remote site. We can see that the first term of Eq. \ref{eq7} is independent of the frequency source at the remote site. In comparison with conventional approaches, this represents another improvement \cite{schediwy2013high, wu2016coherence}.


\section{Delay-limited phase noise PSD}

\begin{figure}[htbp]
\centering
\includegraphics[width=0.85\linewidth]{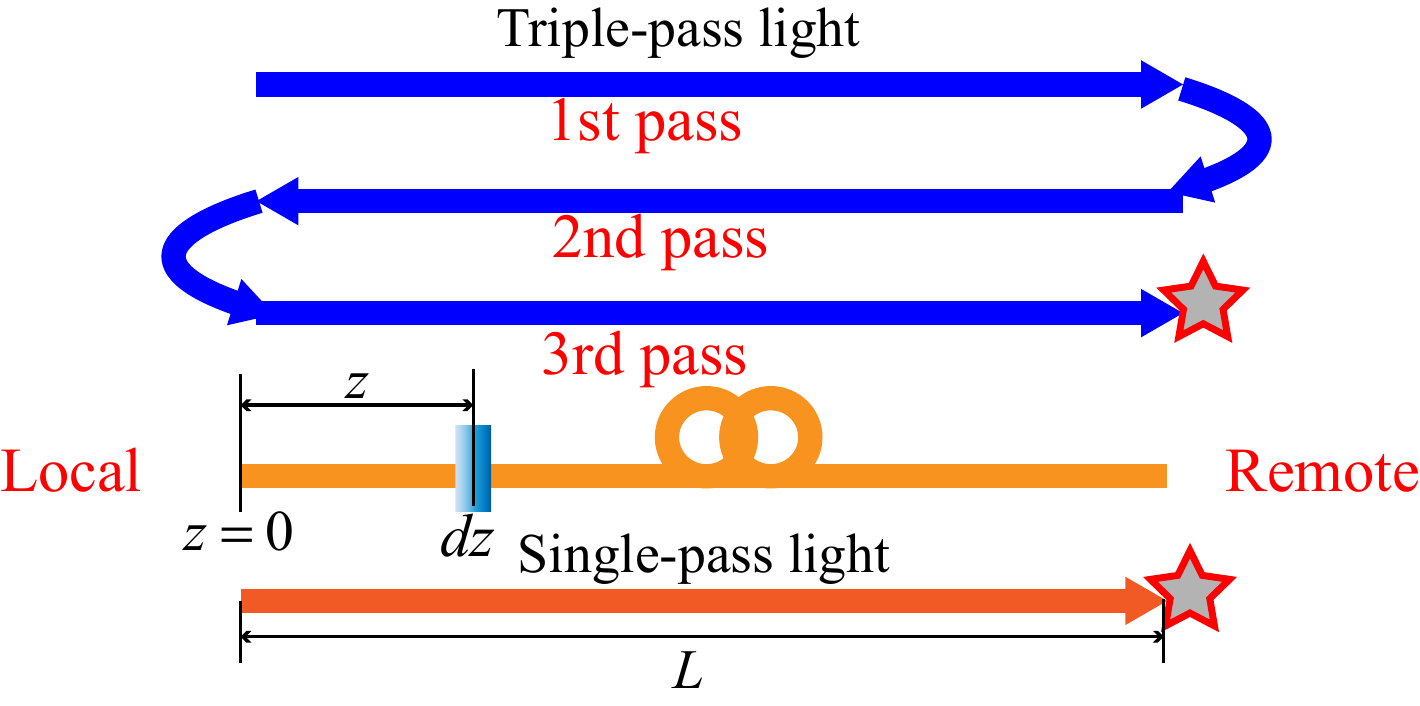}
\caption{Principle of optical frequency through the fiber based on the triple-pass transfer. Here $L$ is the fiber length, and $z$ is the distance between the local end and the site where the fiber phase noise occurs. The phase noise is detected at the remote end at time $t$ by comparing the single-pass light with the triple-pass light.}
\label{fig1_1}
\end{figure}

In this section, we concentrate on theoretical analysis of the delay effect on fiber phase noise compensation in our proposed scheme by using the methods adopted in \cite{williams2008high, bercy2014line}. The delay effect mainly derives from the fiber delay.  As for the triple-pass scheme illustrated in Fig. \ref{fig1_1}, the light is coupled into the fiber at the local end ($z=0$) and is transferred through the fiber for a distance of $L$. The fiber phase noise is collected at the remote end ($z=L$) at time $t=0$. Thus the phase noise at the position $z$ and at the time $t$ can be expressed as $\delta\phi(x,t)$. The fiber delay for a single pass is $\tau_0=L/c_n$, where $c_n$ is the speed of light in the fiber. For the single-pass light, the accumulated fiber phase noise is 
\begin{equation}
\phi_{s}(t)=\int_0^L\delta\phi[z,t-(\tau_0-z/c_n)]dz.
\end{equation}

The Fourier transforms of the above equation is
\begin{equation}
\begin{split}
\widetilde{\phi}_{s}(\omega)&=\int_0^{L}\exp(-i\omega ({\tau_0}-z/c_{n}))\delta\widetilde{\phi}(z,\omega)dz.\\
\end{split}
\label{eq10}
\end{equation}

Similarly, for the triple-pass light, the total fiber phase noise $\phi_t(t)$ collected by accumulating the fiber noise in the first, second and third passes has an expression of 
\begin{widetext}
\begin{align}
\phi_{t}(t)=\int_0^L\delta\phi[x,t-(3\tau_0-x/c_n)]dx+\int_0^L\delta\phi[x,t-(\tau_0+x/c_n)]dx+\int_0^L\delta\phi[x,t-(\tau_0-x/c_n)]dx
\end{align}
\end{widetext}

The beat signal between the single-pass light and the triple-pass light is obtained by comparing $\phi_s(t)$ with $\phi_{t}(t)$. It is obvious that the fiber noise of the single-pass light is equal to that from the third pass of the triple-pass light. Thus they cancel each other out, resulting in the fiber noise of the first and second passes in the beat signal. An error signal $\phi_{\text{error}}(t)$ indicating the fiber phase noise can be expressed as
\begin{equation}
\begin{split}
\phi_{\text{error}}(t)=\int_0^L[&\delta\phi(z,t-(3\tau_0-z/c_n))\\
&+\delta\phi(z,t-(\tau_0+z/c_n))]dz,
\end{split}
\end{equation}

Assuming that the noise is uncorrelated with position, the phase-noise PSD for the fiber-induced phase noise on the one-way light is
\begin{equation}
S_{s}(\omega)=\langle|\widetilde{\phi}_s(\omega)|^2\rangle=\int_0^L\langle|\delta\widetilde{\phi}_s(z, \omega)|^2\rangle dz,
\end{equation}

We divide the angular frequency of the beat signal with a factor of 2 and feed the divided frequency onto the RF port of AOM$_{r2}$. Thus the remaining phase noise in the optical signal can be obtained by subtracting the extracted phase noise $\phi_{\text{error}}$ from the uncompensated single-pass light 

\begin{equation}
\widetilde\phi_{\text{remote}}(\omega)=\widetilde\phi_{s}(\omega)-\frac{1}{2}\widetilde\phi_{\text{error}}(\omega)
\label{eq14}
\end{equation}

With an assistance of Eq. \ref{eq10}, Eq. \ref{eq14} becomes
\begin{equation}
\begin{split}
\widetilde{\phi}_{\text{remote}}(\omega)&=\int_0^{L}\exp(-i\omega (\tau_0-zc_{n}^{-1}))\delta\widetilde{\phi}(z,\omega)dz\\
-\int_{0}^{L}\exp&(-2i\omega\tau_0)\cos(\omega(\tau_0-z/c_n))\delta\widetilde{\phi}(z,\omega)dz
\end{split}
\end{equation}

Again assuming  spatially uncorrelated noise along the fiber, we can have $\langle|\delta\widetilde{\phi}(z, \omega)|^2\rangle=dS_{\text{fiber}}(\omega,z)=S_{\text{fiber}}(\omega)/L$.  The remaining phase noise PSD at the remote site is then 

\begin{widetext}
\begin{align}
S_{\text{remote}}(\omega)&=\bigg \langle\bigg|\int_0^{L}\delta\widetilde{\phi}(z,\omega)[\exp(-i\omega\tau_0)(\exp(i\omega z/c_n)-\exp(-i\omega \tau_0)\cos(\omega(\tau_0-z/c_n)))]dz\bigg|^2\bigg \rangle\\
&\simeq \left(\frac{3}{2}-\cos(2\omega\tau_0)-\frac{1}{2}\text{sinc}(2\omega\tau_0)\right)S_{s}(\omega)\nonumber
\end{align}
\label{eq016}
\end{widetext}

Trigonometric function can be expanded with the Taylor expansion when $\omega\tau_0\ll 1$,  so $\cos(x)\simeq1-x^2/2+\mathcal{O}(x)^4$ and $\text{sinc}(x)\simeq 1-x^2/6+\mathcal{O}(x)^4$. By inserting the expansion results into Eq. 15, then the remaining phase noise PSD at the remote site yields
\begin{equation}
S_{\text{remote}}(\omega)\simeq\left(\frac{7}{3}(\omega\tau_0)^2+\mathcal{O}(\omega\tau_0)^3\right)S_{s}(\omega)
\end{equation}
Once we ignore the high order term of $\mathcal{O}(\omega\tau_0)^3$, the phase noise spectral density has the same expression of
\begin{equation}
S_{\text{remote}}(\omega)\simeq\frac{7}{3}(\omega\tau_0)^2S_{s}(\omega)
\label{eq18}
\end{equation}

The result indicates that the fiber phase noise cannot be thoroughly compensated because of the delay effect and it is proportional to $\tau^2_{0}$ and the uncompensated single-pass fiber phase noise PSD $S_{s}(\omega)$. The residual fiber phase noise PSD in the triple-pass scheme turns out to be a slight worse (a factor of 7) than that of in the conventional double-pass scheme \cite{williams2008high, bercy2014line}.

\section{Experimental apparatus and results}


\subsection{Experimental Apparatus}

We have demonstrated this technique by using the simplest configuration that possesses all the critical elements for optical frequency over a star topology fiber network. The proposed scheme was tested using a narrow-linewidth optical source (NKT X15) at a frequency near 194.3 THz. The signal was then split equally into two parts, one part being transmitted along 145 km spooled fiber link to the remote site 1. Due to the limited available components, here we only test the remote site 1. Two home-made bidirectional Erbium-doped fiber amplifiers (EDFAs) have been adopted after 70 km and 145 km fiber link for  boosting the fading optical signal. This proof-of-principle demonstration was over 145 km, but the technique is equally applicable to longer fiber by inserting more bidirectional amplifiers along the fiber link \cite{droste2013optical2, calonico2014high}. The angular frequencies of AOMs at the local and  remote sites are $\omega_{l}=2\pi\times45$ MHz (AOM$_l$, downshifted mode), $\omega_{r1}=2\pi\times80$ MHz (AOM$_{r1}$, upshifted mode) and $\omega_{r2}=2\pi\times80$ MHz (AOM$_{r2}$, downshifted mode), respectively. To match the driving frequency of AOM$_{r2}$, we mix $E_4$ with a RF signal with a frequency of 45 MHz and the upper sideband is extracted. With this configuration, we can acquire an out-loop heterodyne beat frequency $45$ MHz.

To measure the phase noise of the optical carrier frequency at the remoter, we perform the measurement by feeding the heterodyne beat frequency together with a stable frequency reference to a phase detector. The voltage fluctuations at the phase detector output are then measured with an FFT-analyzer to obtain the phase fluctuations $S_{\phi}(\omega)$. Here we measure the phase noise power spectral density (PSD) for the stabilized fiber link and for the free-running link where no noise cancellation is applied. Additionally, to effectively measure the transfer stability to the remote site, we use a frequency counter, which is referenced to the frequency source at the local site, to record the out-loop heterodyne beat frequency.


\subsection{Interferometer noise floor}

\begin{figure}[htbp]
\centering
\includegraphics[width=1\linewidth]{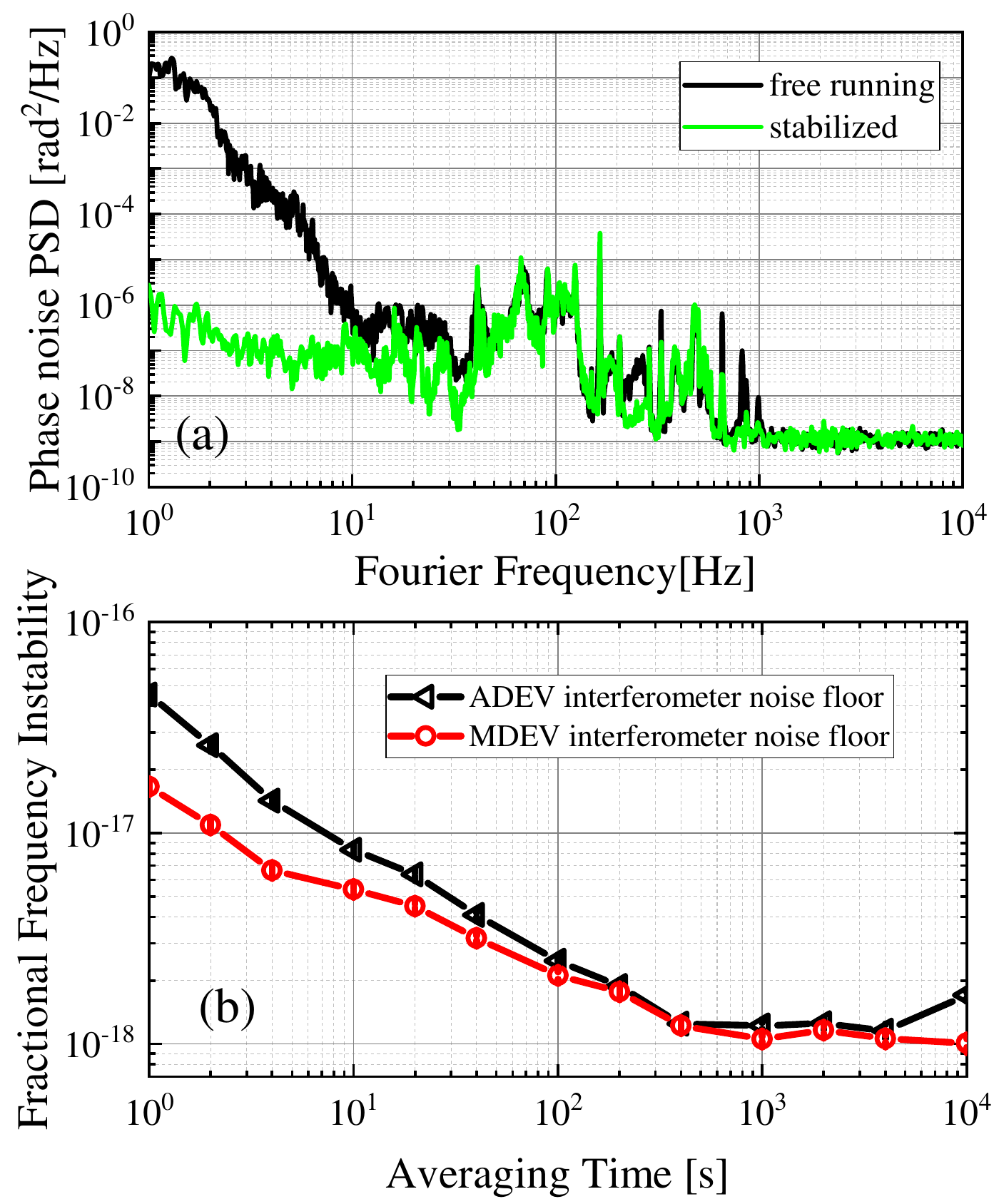}
\caption{Interferometer noise floor measurement results. (a) Measured phase noise power spectral density (PSD) of a free-running interferometer (black curves) and stabilized interferometer (green curves). (b)  Black left triangles and red circles are measured stabilized interferometer derived from a nonaveraging ($\Pi$-type) frequency counter in terms of ADEV and  an averaging ($\Lambda$-type) frequency counter in terms of MDEV. The error bars are inside of the symbols. }
\label{fig2}
\end{figure}

Figure \ref{fig2} (a) shows that a wide spread frequency domain characterization of phase instability is the PSD of phase fluctuations $S_{\phi}(f)=\langle|\tilde{\phi}(\omega)|^2\rangle$ for the out-of-loop beat signal of the unstabilized and stabilized interferometer, respectively.  The noise floor  was measured by optically short-cutting the stabilized link by connecting the output of the local acousto-optic modulator to the input of the remote EDFA using an optical attenuator. The phase noise of the unstabilized interferometer (black curve in Fig. \ref{fig2} (a)) is dominated by $S_{\phi}(f)\sim10^{-1}/f^5$ rad$^2$/Hz up to a Fourier frequency of $f=10$ Hz while above 10 Hz a white phase noise level of  $S_{\phi}(f)=h_0/f^0$ rad$^2$/Hz is reached.  Once the interferometer is stabilized by means of the proposed passive phase noise cancellation, the fluctuations of the path length and those arising from devices that are inside of the loop are corrected. The phase noise is suppressed by approximately 50 dB at 1 Hz. The stabilized interferometer is dominated by a white phase noise level of  $S_{\phi}(f)\sim10^{-7}/f^0$ rad$^2$/Hz. Figure \ref{fig2} (b) illustrates the corresponding fractional frequency instability. Measurements of the frequency instability for longer averaging times calculated by overlapping Allan deviation (ADEV) (triangles) and modified Allan deviation (MDEV) (circles) taken with the counters operating in $\Pi$-averaging mode and $\Lambda$-averaging mode, respectively, with 1 s gate time \cite{dawkins2007considerations}.  We can clearly see that the variable noise floor, which for averaging times beyond around 300 s is approximately $10^{-18}$. We attribute the variation in the $10^{-18}$ range to differential temperature fluctuations in the local and the remote optical setup and received RF power fluctuations introduced by the polarization variations in the fiber link \cite{williams2008high}. Therefore, we regard the observed  instability around $10^{-18}$ as fortuitous in our present setup, where the temperature of the optical setups is not controlled passively or actively.  To further determine the electronic noise limit of our measurement capabilities, the electrical path was locked to an RF synthesizer and its frequency was counted. A fractional frequency instability in terms of ADEV $1.6\times10^{-17}/\tau$ was achieved which includes electronic noise due to the RF components and the frequency counting system.

\subsection{Main fiber link}

\begin{figure}[htbp]
\centering
\includegraphics[width=0.95\linewidth]{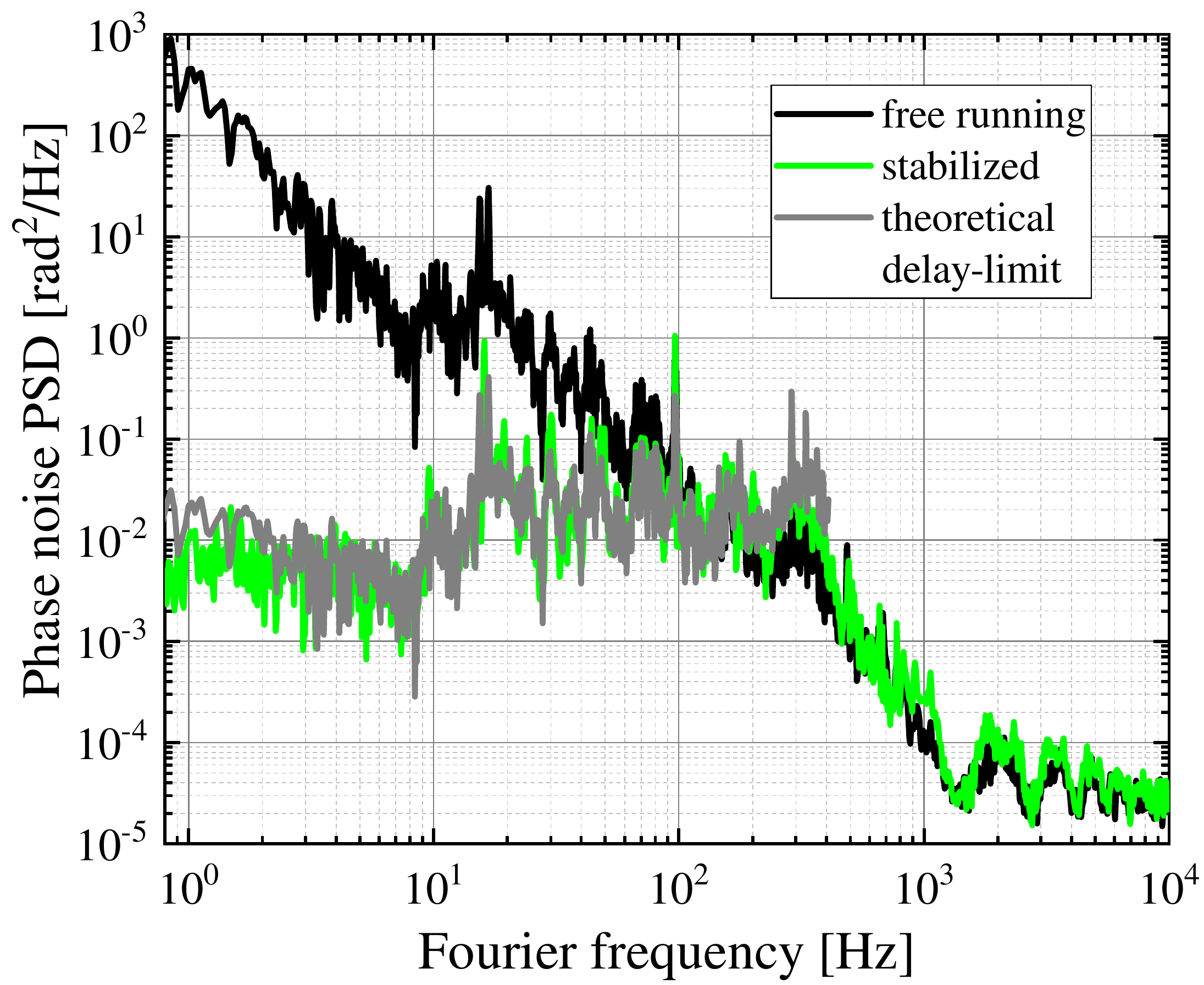}
\caption{Measured phase noise power spectral density (PSD) of a 145 km free-running (black curve) and stabilized link with fiber noise cancellation (green curve) for the stabilized optical frequency. The grey curve is the theoretical prediction by using Eq. \ref{eq18}.  }
\label{fig4}
\end{figure}

Figure \ref{fig4} shows the phase noise PSD for the 145 km stabilized fiber link (green curve), and for the 145 km free-running link (black curve), where no phase noise cancellation was applied.  We find that unlocked phase noise on our fiber link approximately follows a power-law dependence, $S_{\phi}(f)\sim h_2/f^2$ rad$^2$/Hz, for $f<1$ kHz, indicating that white frequency noise is dominating in the free running fiber link, where $h_2\sim430$ is a coefficient that varies for different fiber links. Once the fiber noise cancellation system at the remote end is applied, the phase noise PSD is remarkably suppressed down to $\sim10^{-2}$ rad$^2$/Hz within its delay-unsuppressed below $1/4\tau_0\simeq 345$ Hz, showing the remaining noise is mainly determined by the white phase noise. It is important to note that strong servo bumps disappeared. In comparison with conventional  active phase noise cancellation schemes, the disappeared servo bumps represent another improvement because this reduce the integrated phase noise \cite{williams2008high}.  From Fig. \ref{fig4}, we also can find the measured locked phase noise to be in good agreement with the theoretical value (grey curves) as predicted in Eq. \ref{eq18}. This, on other hand, confirms the correctness of the theoretical model presented in Sec. III.

\begin{figure}[htbp]
\centering
\includegraphics[width=0.95\linewidth]{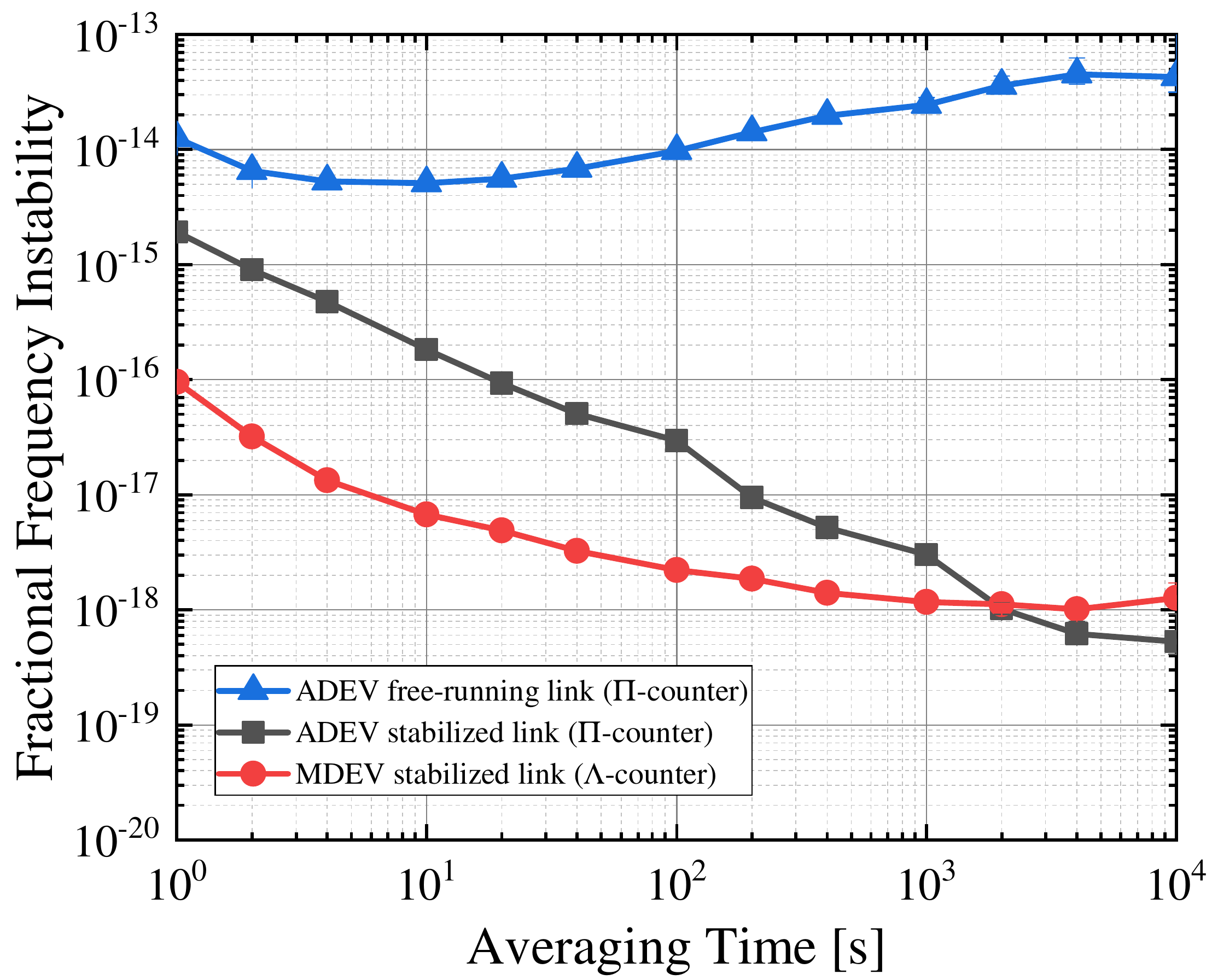}
\caption{Measured fractional frequency instability of the 145-km free-running fiber link (blue up triangles) and the stabilized link (black squares) derived from a nonaveraging ($\Pi$-type) frequency counter and expressed as the ADEV. To obtain a significantly shorter measurement time and to distinguish between white phase, flicker phase, and other noise types, averaging (overlapping $\Lambda$-type) frequency counters is used for which the MDEV (red circles) measures the instability. The error bars are inside of the symbols. }
\label{fig3}
\end{figure}

Complementary to the frequency-domain characterization, a time-domain feature of the fractional frequency instability is performed.  Figure \ref{fig3} shows the instability of the free-running link (up triangles). It fluctuates in the $10^{-14}$ range with an overall mean of the fractional frequency offset of here $\sim2\times10^{-14}$ . The instability indicates processes with several periods of temperature fluctuations.  In case of ADEV, it is around $1.9\times10^{-15}$ at an averaging time of 1 s and reaches a floor in the $10^{-18}$-range for a few thousands of seconds of averaging time.  The filled circles represent the result of applying the MDEV formalism to the $\Lambda$-type data, The $\Lambda$-MDEVs fall off as $9.6\times10^{-17}$$/\tau^2$ for averaging times up to a few seconds as expected e.g. for white phase noise. For the measurement, a flicker frequency noise floor around $10^{-18}$ seems to be reached after about 1,000 s of averaging time $\tau$.  This is to be expected from the variations of the current noise floor of the setup without the long fiber link as shown in Fig. \ref{fig2} (b). 

\begin{figure}[htbp]
\centering
\includegraphics[width=0.99\linewidth]{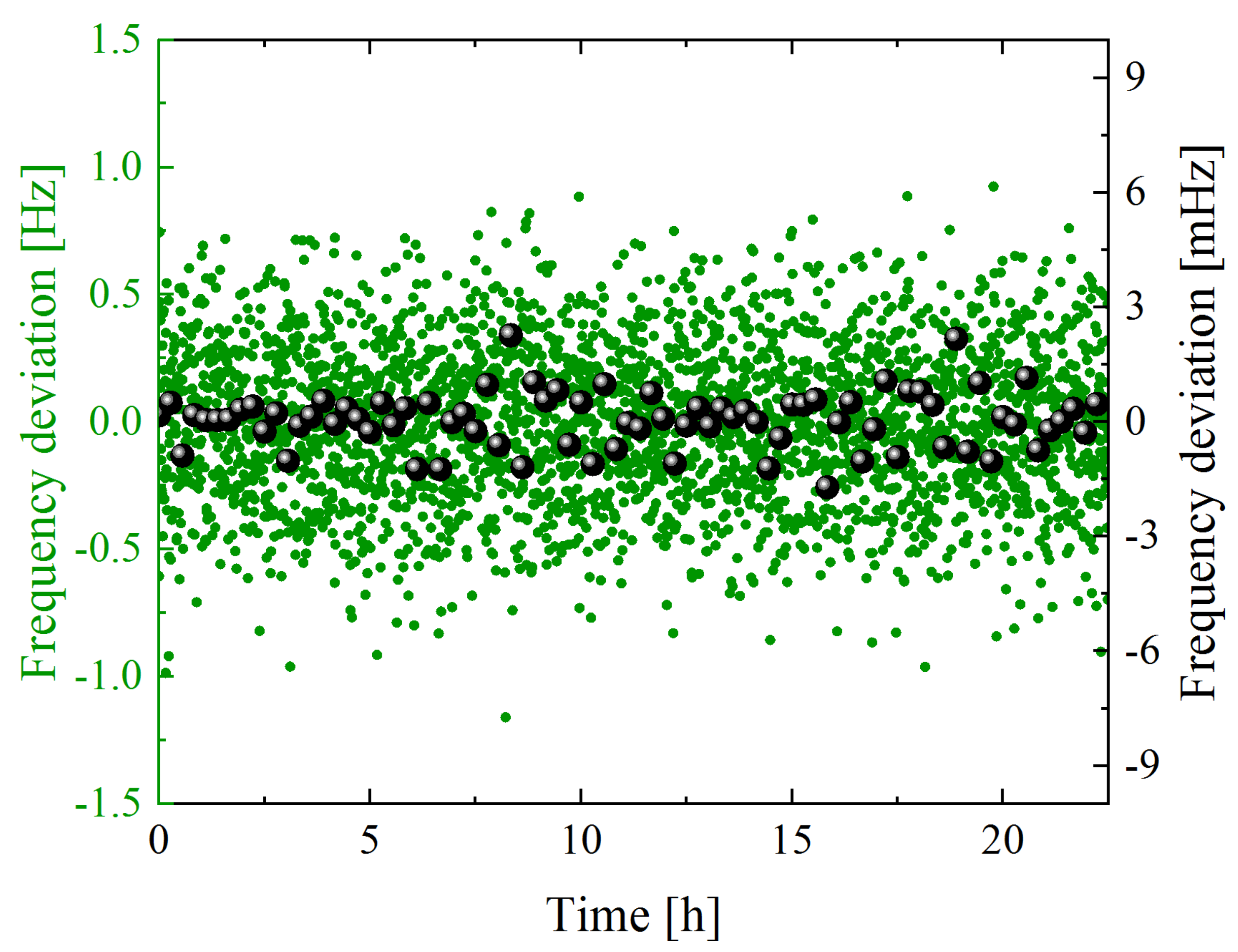}
\caption{Frequency deviation of the transferred light from the expected value when transferring through 145-km fiber link (green dots). The relative frequency deviation is measured by $\Pi$-type frequency counter. The arithmetic means of all cycle-slip free 1,000 s intervals have been computed. From the resulting 81 data points (black dots, right frequency axis), a fractional difference between sent and transferred frequency of $(-0.36\pm2.6)\times10^{-18}$ is calculated.}
\label{fig5}
\end{figure}

Figure \ref{fig5} displays the frequency deviation of the transferred light from the expected value. Here we recorded the relevant beat notes with 1 s gate time and with $\Pi$-type counter, to avoid overweighting the center parts of the data sets with the triangular weighting of the $\Lambda$-type counter. The frequency the beat signal was measured in 81,000 s on a counter with a gate time of 1 s.  The distribution of the frequency deviation over counts follows a Gaussian function. We divided the data set into 81 subsets with a length of continuous 1,000 s time period. The mean value of each subset is calculated, as shown with black filled circles in Fig. \ref{fig5}. The 81 data points have an arithmetic mean of $-68.7$ $\mu$Hz (-3.6$\times10^{-19}$) and a standard deviation of $494$ $\mu$Hz (2.6$\times10^{-18}$). Consequently, the statistical fractional uncertainty of the 81,000 data points is calculated to be $(-0.36\pm2.6)\times10^{-18}$. These results indicate that the system meets the requirements of remote comparisons of optical atomic clocks \cite{mcgrew2019towards, lisdat2016clock, grotti2018geodesy}.


To note that in the current setup we only stabilize the fiber path starting at the input of AOM at the local site and ending at the FM at the remote site, but thermal and acoustic perturbations affecting all other fibers or fiber components involved in the optical carrier transfer path. In particular, the interferometer was built with a fiber-coupled AOM (AOM$_{r2}$), which caused non-optimal spatial design and thus involved relatively long uncompensated fibers and thermal effect. In the immediate next step, we are also planning to stabilize the uncompensated path with passive or active methods \cite{predehl2012920, droste2013optical2, droste2014optical}.


Because of the similarity among the test results of different remote site, we just show the test results of a typical remote sites in a star topology fiber network. As discussed in \cite{schediwy2013high, wu2016coherence}, our scheme, in principle, can support multiple users simultaneously by choosing different AOM-driving frequencies in different channels to make the beat notes at different remote users shift by a few megahertz. Then in each channel, the remote user uses a narrow bandwidth band pass filter to filter out the signal for noise compensation. The number of remote users for this scheme will be determined by some technical problems such as the bandwidth of the band pass filters. Additionally,  there is 3 dB insertion loss by adding one more remote site, proper EDFAs and electrical amplifiers can be used to amplify the desired optical signals and detected RF signals. Thus, it ensures that multiple remote sites can be inserted in a star topology fiber network 

\section{Conclusion}

In summary, we have presented a new method, making a stable optical frequency available at remote user sites. The phase drift generated by fiber-length variations can be automatically eliminated by optical frequency mixing and shifting at the remote sites with assistance of AOMs. Neither phase discrimination nor active phase tracking is required due to the open-loop design, mitigating some technical difficulties in conventional active phase noise cancellation. Moreover, our approach, in principle, doesn't need the remote signal source. Consequently, the signal source with a high frequency stability, is no longer needed, which simplifies the system and makes the system cost-effective. We experimentally demonstrate the transfer of optical frequency to one remote site through 145 km, lab-based spooled fiber links. After being compensated, delivering an optical frequency to multi-node with the relative frequency instability in terms of ADEV measured by $\Pi$-mode frequency counter is $1.9\times10^{-15}$ at 1 s and $ 5.3\times10^{-19}$ at 10,000 s. The frequency uncertainty of the light after transferring through the fiber relative to that of the input light is $(-0.36\pm2.6)\times10^{-18}$ for the 145 km fiber link. 

The proposed technique considerably simplifies future efforts to make coherent optical frequency signals available to many users, enabling the above mentioned applications, and paving a path towards a redefinition of the unit of time, the SI second through regular and practical international comparisons of optical clocks \cite{mcgrew2019towards, oelker2019demonstration, hachisu2018months, yao2017incorporating, Grebing:16}.

\ifCLASSOPTIONcaptionsoff
  \newpage
\fi



%
\bibliographystyle{IEEEtran}
\bibliography{Optics}

\begin{thebibliography}{10}
\providecommand{\url}[1]{#1}
\csname url@samestyle\endcsname
\providecommand{\newblock}{\relax}
\providecommand{\bibinfo}[2]{#2}
\providecommand{\BIBentrySTDinterwordspacing}{\spaceskip=0pt\relax}
\providecommand{\BIBentryALTinterwordstretchfactor}{4}
\providecommand{\BIBentryALTinterwordspacing}{\spaceskip=\fontdimen2\font plus
\BIBentryALTinterwordstretchfactor\fontdimen3\font minus
  \fontdimen4\font\relax}
\providecommand{\BIBforeignlanguage}[2]{{%
\expandafter\ifx\csname l@#1\endcsname\relax
\typeout{** WARNING: IEEEtran.bst: No hyphenation pattern has been}%
\typeout{** loaded for the language `#1'. Using the pattern for}%
\typeout{** the default language instead.}%
\else
\language=\csname l@#1\endcsname
\fi
#2}}
\providecommand{\BIBdecl}{\relax}
\BIBdecl

\bibitem{ludlow2015optical}
A.~D. Ludlow, M.~M. Boyd, J.~Ye, E.~Peik, and P.~O. Schmidt, ``Optical atomic
  clocks,'' \emph{Reviews of Modern Physics}, vol.~87, no.~2, p. 637, 2015.

\bibitem{poli2014optical}
N.~Poli, C.~Oates, P.~Gill, and G.~Tino, ``Optical atomic clocks,'' \emph{arXiv
  preprint arXiv:1401.2378}, 2014.

\bibitem{PhysRevLett.120.103201}
G.~E. Marti, R.~B. Hutson, A.~Goban, S.~L. Campbell, N.~Poli, and J.~Ye,
  ``Imaging optical frequencies with $100\text{ }\text{
  }\ensuremath{\mu}\mathrm{Hz}$ precision and $1.1\text{ }\text{
  }\ensuremath{\mu}\mathrm{m}$ resolution,'' \emph{Physical review letters},
  vol. 120, p. 103201, 2018.

\bibitem{Schioppo:2016aa}
M.~Schioppo, R.~C. Brown, W.~F. McGrew, N.~Hinkley, R.~J. Fasano, K.~Beloy,
  T.~H. Yoon, G.~Milani, D.~Nicolodi, J.~A. Sherman, N.~B. Phillips, C.~W.
  Oates, and A.~D. Ludlow, ``Ultrastable optical clock with two cold-atom
  ensembles,'' \emph{Nature Photonics}, vol.~11, pp. 48--52, 2016.

\bibitem{McGrew:2018aa}
W.~F. McGrew, X.~Zhang, R.~J. Fasano, S.~A. Sch{\"a}ffer, K.~Beloy,
  D.~Nicolodi, R.~C. Brown, N.~Hinkley, G.~Milani, M.~Schioppo, T.~H. Yoon, and
  A.~D. Ludlow, ``Atomic clock performance enabling geodesy below the
  centimetre level,'' \emph{Nature}, vol. 564, no. 7734, pp. 87--90, 2018.

\bibitem{Giorgetta:2013aa}
F.~R. Giorgetta, W.~C. Swann, L.~C. Sinclair, E.~Baumann, I.~Coddington, and
  N.~R. Newbury, ``Optical two-way time and frequency transfer over free
  space,'' \emph{Nature Photonics}, vol.~7, pp. 434--438, 2013.

\bibitem{Riehle:2017aa}
F.~Riehle, ``Optical clock networks,'' \emph{Nature Photonics}, vol.~11, pp. 25
  EP --, 01 2017.

\bibitem{ma1994delivering}
L.-S. Ma, P.~Jungner, J.~Ye, and J.~L. Hall, ``Delivering the same optical
  frequency at two places: accurate cancellation of phase noise introduced by
  an optical fiber or other time-varying path,'' \emph{Optics letters},
  vol.~19, no.~21, pp. 1777--1779, 1994.

\bibitem{foreman2007coherent}
S.~M. Foreman, A.~D. Ludlow, M.~H. De~Miranda, J.~E. Stalnaker, S.~A. Diddams,
  and J.~Ye, ``Coherent optical phase transfer over a 32-km fiber with 1 s
  instability at $10^{-17}$,'' \emph{Physical review letters}, vol.~99, no.~15,
  p. 153601, 2007.

\bibitem{daussy2005long}
C.~Daussy, O.~Lopez, A.~Amy-Klein, A.~Goncharov, M.~Guinet, C.~Chardonnet,
  F.~Narbonneau, M.~Lours, D.~Chambon, S.~Bize \emph{et~al.}, ``Long-distance
  frequency dissemination with a resolution of $10^{-17}$,'' \emph{Physical
  review letters}, vol.~94, no.~20, p. 203904, 2005.

\bibitem{droste2013optical2}
S.~Droste, F.~Ozimek, T.~Udem, K.~Predehl, T.~H{\"a}nsch, H.~Schnatz,
  G.~Grosche, and R.~Holzwarth, ``Optical-frequency transfer over a single-span
  1840 km fiber link,'' \emph{Physical review letters}, vol. 111, no.~11, p.
  110801, 2013.

\bibitem{calonico2014high}
D.~Calonico, E.~Bertacco, C.~Calosso, C.~Clivati, G.~Costanzo, M.~Frittelli,
  A.~Godone, A.~Mura, N.~Poli, D.~Sutyrin \emph{et~al.}, ``High-accuracy
  coherent optical frequency transfer over a doubled 642-km fiber link,''
  \emph{Applied Physics B}, vol. 117, no.~3, pp. 979--986, 2014.

\bibitem{clivati2017vlbi}
C.~Clivati, R.~Ambrosini, T.~Artz, A.~Bertarini, C.~Bortolotti, M.~Frittelli,
  F.~Levi, A.~Mura, G.~Maccaferri, M.~Nanni \emph{et~al.}, ``A {VLBI}
  experiment using a remote atomic clock via a coherent fibre link,''
  \emph{Scientific reports}, vol.~7, p. 40992, 2017.

\bibitem{wang2015square}
B.~Wang, X.~Zhu, C.~Gao, Y.~Bai, J.~Dong, and L.~Wang, ``Square kilometre array
  telescope---precision reference frequency synchronisation via 1f-2f
  dissemination,'' \emph{Scientific reports}, vol.~5, p. 13851, 2015.

\bibitem{lisdat2016clock}
C.~Lisdat, G.~Grosche, N.~Quintin, C.~Shi, S.~Raupach, C.~Grebing, D.~Nicolodi,
  F.~Stefani, A.~Al-Masoudi, S.~D{\"o}rscher \emph{et~al.}, ``A clock network
  for geodesy and fundamental science,'' \emph{Nature communications}, vol.~7,
  p. 12443, 2016.

\bibitem{hu2017atom}
L.~Hu, N.~Poli, L.~Salvi, and G.~M. Tino, ``Atom interferometry with the {S}r
  optical clock transition,'' \emph{Physical review letters}, vol. 119, no.~26,
  p. 263601, 2017.

\bibitem{grotti2018geodesy}
J.~Grotti, S.~Koller, S.~Vogt, S.~H{\"a}fner, U.~Sterr, C.~Lisdat, H.~Denker,
  C.~Voigt, L.~Timmen, A.~Rolland \emph{et~al.}, ``Geodesy and metrology with a
  transportable optical clock,'' \emph{Nature Physics}, p.~1, 2018.

\bibitem{grosche2014eavesdropping}
G.~Grosche, ``Eavesdropping time and frequency: phase noise cancellation along
  a time-varying path, such as an optical fiber,'' \emph{Optics letters},
  vol.~39, no.~9, pp. 2545--2548, 2014.

\bibitem{bai2013fiber}
Y.~Bai, B.~Wang, X.~Zhu, C.~Gao, J.~Miao, and L.~Wang, ``Fiber-based
  multiple-access optical frequency dissemination,'' \emph{Optics letters},
  vol.~38, no.~17, pp. 3333--3335, 2013.

\bibitem{bercy2014line}
A.~Bercy, S.~Guellati-Khelifa, F.~Stefani, G.~Santarelli, C.~Chardonnet, P.-E.
  Pottie, O.~Lopez, and A.~Amy-Klein, ``In-line extraction of an ultrastable
  frequency signal over an optical fiber link,'' \emph{Journal of the Optical
  Society of America B}, vol.~31, no.~4, pp. 678--685, 2014.

\bibitem{schediwy2013high}
S.~W. Schediwy, D.~Gozzard, K.~G. Baldwin, B.~J. Orr, R.~B. Warrington,
  G.~Aben, and A.~N. Luiten, ``High-precision optical-frequency dissemination
  on branching optical-fiber networks,'' \emph{Optics letters}, vol.~38,
  no.~15, pp. 2893--2896, 2013.

\bibitem{wu2016coherence}
L.~Wu, Y.~Jiang, C.~Ma, H.~Yu, Z.~Bi, and L.~Ma, ``Coherence transfer of
  subhertz-linewidth laser light via an optical fiber noise compensated by
  remote users,'' \emph{Optics letters}, vol.~41, no.~18, pp. 4368--4371, 2016.

\bibitem{falke2012delivering}
S.~Falke, M.~Misera, U.~Sterr, and C.~Lisdat, ``Delivering pulsed and phase
  stable light to atoms of an optical clock,'' \emph{Applied Physics B}, vol.
  107, no.~2, pp. 301--311, 2012.

\bibitem{gozzard2018stabilized}
D.~Gozzard, S.~Schediwy, B.~Stone, M.~Messineo, and M.~Tobar, ``Stabilized
  free-space optical frequency transfer,'' \emph{Physical Review Applied},
  vol.~10, no.~2, p. 024046, 2018.

\bibitem{pan2016passive}
S.~Pan, J.~Wei, and F.~Zhang, ``Passive phase correction for stable radio
  frequency transfer via optical fiber,'' \emph{Photonic Network
  Communications}, vol.~31, no.~2, pp. 327--335, 2016.

\bibitem{he2013stable}
Y.~He, B.~J. Orr, K.~G. Baldwin, M.~J. Wouters, A.~N. Luiten, G.~Aben, and
  R.~B. Warrington, ``Stable radio-frequency transfer over optical fiber by
  phase-conjugate frequency mixing,'' \emph{Optics express}, vol.~21, no.~16,
  pp. 18\,754--18\,764, 2013.

\bibitem{huang2016fiber}
R.~Huang, G.~Wu, H.~Li, and J.~Chen, ``Fiber-optic radio frequency transfer
  based on passive phase noise compensation with frequency dividing and
  filtering,'' \emph{Optics letters}, vol.~41, no.~3, pp. 626--629, 2016.

\bibitem{li2014phase}
D.~Li, D.~Hou, E.~Hu, and J.~Zhao, ``Phase conjugation frequency dissemination
  based on harmonics of optical comb at 10- 17 instability level,''
  \emph{Optics letters}, vol.~39, no.~17, pp. 5058--5061, 2014.

\bibitem{yu2014stable}
L.~Yu, R.~Wang, L.~Lu, Y.~Zhu, C.~Wu, B.~Zhang, and P.~Wang, ``Stable radio
  frequency dissemination by simple hybrid frequency modulation scheme,''
  \emph{Optics letters}, vol.~39, no.~18, pp. 5255--5258, 2014.

\bibitem{wu2013stable}
Z.~Wu, Y.~Dai, F.~Yin, K.~Xu, J.~Li, and J.~Lin, ``Stable radio frequency phase
  delivery by rapid and endless post error cancellation,'' \emph{Optics
  letters}, vol.~38, no.~7, pp. 1098--1100, 2013.

\bibitem{ma2018delay}
C.-q. Ma, L.-F. Wu, J.~Gu, Y.-H. Chen, and G.-Q. Chen, ``Delay effect on
  coherent transfer of optical frequency based on a triple-pass scheme,''
  \emph{Chinese Physics Letters}, vol.~35, no.~8, p. 080601, 2018.

\bibitem{williams2008high}
P.~A. Williams, W.~C. Swann, and N.~R. Newbury, ``High-stability transfer of an
  optical frequency over long fiber-optic links,'' \emph{Journal of the Optical
  Society of America B}, vol.~25, no.~8, pp. 1284--1293, 2008.

\bibitem{dawkins2007considerations}
S.~T. Dawkins, J.~J. McFerran, and A.~N. Luiten, ``Considerations on the
  measurement of the stability of oscillators with frequency counters,''
  \emph{Ieee transactions on ultrasonics, ferroelectrics, and frequency
  control}, vol.~54, no.~5, pp. 918--925, 2007.

\bibitem{mcgrew2019towards}
W.~F. McGrew, X.~Zhang, H.~Leopardi, R.~Fasano, D.~Nicolodi, K.~Beloy, J.~Yao,
  J.~A. Sherman, S.~A. Schaeffer, J.~Savory \emph{et~al.}, ``Towards the
  optical second: verifying optical clocks at the si limit,'' \emph{Optica},
  vol.~6, no.~4, pp. 448--454, 2019.

\bibitem{predehl2012920}
K.~Predehl, G.~Grosche, S.~Raupach, S.~Droste, O.~Terra, J.~Alnis, T.~Legero,
  T.~H{\"a}nsch, T.~Udem, R.~Holzwarth \emph{et~al.}, ``A 920-kilometer optical
  fiber link for frequency metrology at the 19th decimal place,''
  \emph{Science}, vol. 336, no. 6080, pp. 441--444, 2012.

\bibitem{droste2014optical}
S.~Droste, K.~Predehl, T.~H{\"a}nsch, T.~Udem, R.~Holzwarth, F.~Ozimek,
  H.~Schnatz, and G.~Grosche, ``Optical frequency transfer via 1840 km fiber
  link with superior stability,'' in \emph{CLEO: Science and
  Innovations}.\hskip 1em plus 0.5em minus 0.4em\relax Optical Society of
  America, 2014, pp. SW3O--3.

\bibitem{oelker2019demonstration}
E.~Oelker, R.~Hutson, C.~Kennedy, L.~Sonderhouse, T.~Bothwell, A.~Goban,
  D.~Kedar, C.~Sanner, J.~Robinson, G.~Marti \emph{et~al.}, ``Demonstration of
  4.8$\times 10^{-17}$ stability at 1 s for two independent optical clocks,''
  \emph{Nature Photonics}, vol.~13, no.~10, pp. 714--719, 2019.

\bibitem{hachisu2018months}
H.~Hachisu, F.~Nakagawa, Y.~Hanado, and T.~Ido, ``Months-long real-time
  generation of a time scale based on an optical clock,'' \emph{Scientific
  reports}, vol.~8, no.~1, p. 4243, 2018.

\bibitem{yao2017incorporating}
J.~Yao, T.~E. Parker, N.~Ashby, and J.~Levine, ``Incorporating an optical clock
  into a time scale,'' \emph{IEEE transactions on ultrasonics, ferroelectrics,
  and frequency control}, vol.~65, no.~1, pp. 127--134, 2017.

\bibitem{Grebing:16}
C.~Grebing, A.~Al-Masoudi, S.~D\"{o}rscher, S.~H\"{a}fner, V.~Gerginov,
  S.~Weyers, B.~Lipphardt, F.~Riehle, U.~Sterr, and C.~Lisdat, ``Realization of
  a timescale with an accurate optical lattice clock,'' \emph{Optica}, vol.~3,
  no.~6, pp. 563--569, 2016.

\end{thebibliography}

\begin{IEEEbiographynophoto}{Liang Hu}
received the B.S. degree from Hangzhou Dianzi University, China, in 2011, and the M.S. degree from Shanghai Jiao Tong University, China, in 2014. He received the Ph.D. degree from University of Florence, Italy, in 2017 during which he was a Marie-Curie Early Stage Researcher at FACT project. He is currently a Tenure-Track Assistant Researcher in the State Key Laboratory of Advanced Optical Communication Systems and Networks, Department of Electronic Engineering, Shanghai Jiao Tong University, China. His current research interests include photonic signal transmission and atom interferometry.
\end{IEEEbiographynophoto}

\begin{IEEEbiographynophoto}{Xueyang Tian}
received the B.S. degree from Shanghai Dianji University, China, in 2017. She is currently a graduate student in the State Key Laboratory of Advanced Optical Communication Systems and Networks, Department of Electronic Engineering, Shanghai Jiao Tong University, China. Her current research interests include photonic signal transmission.
\end{IEEEbiographynophoto}

\begin{IEEEbiographynophoto}{Guiling Wu}
received the B.S. degree from Haer Bing Institute of Technology,
China, in 1995, and the M.S. and Ph.D. degrees from Huazhong University
of Science and Technology, China, in 1998 and 2001, respectively.
He is currently a Professor in the State Key Laboratory of Advanced Optical
Communication Systems and Networks, Department of Electronic Engineering,
Shanghai Jiao Tong University, China. His current research interests include
photonic signal processing and transmission.
\end{IEEEbiographynophoto}

\begin{IEEEbiographynophoto}{Mengya Kong }
received the B.S degree from Nanjing University of Posts and Telecommunications, China, in 2017. She is currently a graduate student in the State Key Laboratory of Advanced Optical Communication Systems and Networks, Department of Electronic Engineering, Shanghai Jiao Tong University. Her main research interests include opto-electronic devices and RF circuit integration, microwave frequency transfer and system applications.
\end{IEEEbiographynophoto}

\begin{IEEEbiographynophoto}{Jianguo Shen}
received his Bachelor Degree on Physics Educations from Zhengjiang Normal University in 2002, Master Degree on Circuit and system from Hangzhou Dianzi University in 2007, Ph.D on Electromagnetic and Microwave Technology from Shanghai Jiaotong University in 2015. Currently, he is an associate professor at Zhejiang normal university of china. His research interests include microwave photonic signals processing and time and frequency transfer over the optical fiber.   
\end{IEEEbiographynophoto}
\newpage
\newpage


\begin{IEEEbiographynophoto}{Jianping Chen}
received the B.S. degree from Zhejiang University, China, in
1983, and the M.S. and Ph.D. degrees from Shanghai Jiao Tong University,
China, in 1986 and 1992, respectively. He is currently a Professor in the State
Key Laboratory of Advanced Optical Communication Systems and Networks,
Department of Electronic Engineering, Shanghai Jiao Tong University. His
main research interests include opto-electronic devices and integration, photonic
signal processing, and system applications. He is a Principal Scientist of National
Basic Research Program of China (also known as 973 Program).
\end{IEEEbiographynophoto}

\end{document}